\documentclass{PoS}

\usepackage{slashed}
\usepackage{multirow}
\usepackage{sidecap}
\usepackage{verbatim}

\title{Discovery potential for $T' \to tZ$ in the trilepton channel at the LHC}

\ShortTitle{$T' \to tZ$ discovery potential at the LHC}

\author{\speaker{Lorenzo Basso}\thanks{The work of LB is supported by the Theorie-LHC France initiative of the CNRS/IN2P3 and by the French ANR 12 JS05 002 01 BATS@LHC.}\\
        Universit\`e de Strasbourg, IPHC, 23 rue du Loess 67037 Strasbourg, France\\
	CNRS, UMR7178, 67037 Strasbourg, France\\
        E-mail: \email{lorenzo.basso@iphc.cnrs.fr}}

\author{Jeremy Andrea\\
        Universit\`e de Strasbourg, IPHC, 23 rue du Loess 67037 Strasbourg, France\\
	CNRS, UMR7178, 67037 Strasbourg, France\\
        E-mail: \email{jeremy.andrea@iphc.cnrs.fr}}

\abstract{The LHC discovery potential of heavy top partners decaying into a top quark and a $Z$ boson is studied in the trilepton channel at 13 TeV in the single production mode. The clean multilepton final state allows to strongly reduce the background contaminations and to reconstruct the $T'$ mass. We show that a simple cut-and-count analysis probes the parameter space of a simplified model as efficiently as a dedicated multivariate analysis. The trilepton signature finally turns out to be able to probe $T'$ masses up to $2.0$ TeV, when mixing with first generation quarks is included. The reinterpretation in terms of the top-$Z$-quark anomalous coupling is shown.}

\FullConference{The European Physical Society Conference on High Energy Physics\\
		22--29 July 2015\\
		Vienna, Austria}

\begin{document}

\section{Introduction}
The primary scope of the LHC Run-II is to further understand the newly discovered Higgs boson and to eventually make new discoveries. 
In all generality, it is very common in beyond the standard model theories that new heavy fermions arise to stabilise the Higgs boson mass and to protect it from dangerous quadratic divergences. In many cases, these new fermions are heavy partners of the third generation quarks with vector-like couplings. They are frequently predicted by many new physics scenarios, including Extra Dimensions, Little Higgs Models, and Composite Higgs Models (for a recent review, see Ref.~\cite{Ellis:2014dza}). The common feature of these heavy quarks is to decay into a standard model quark and a $W^\pm$ boson, a $Z$ bosons, or a Higgs boson. Here we will focus on the case of a singlet heavy quark: the {\it{top partner}} or $T'$. Recent limits from ATLAS and CMS lie within 690-780 GeV~\cite{expVLQ}, depending on the considered final state.

These searches are performed in the QCD-like pair production channel and do not typically consider intergenerational mixing. 
However, the top partners can mix in a sizable way with lighter quarks while remaining compatible with the current experimental constraints~\cite{Buchkremer:2013bha}.
Beside opening up the decay channel into a standard model boson plus a light quark, the mixing with the first generation also enhances the single production, especially due to the presence of valence quarks in the initial state. Even without mixing, the single production cross sections at the upcoming LHC energies become competitive with the pair production ones.
Based on Ref.~\cite{Basso:2014apa}, here we summarise the study of the LHC discovery potential of the $T' \to tZ$ channel in the trilepton decay mode in single production at $\sqrt{s}=13$ TeV, for a singlet $T'$ quark mixing with the first generation.  To capture all the essential features of the new heavy top quark while remaining as model independent as possible, the study here presented is performed in the framework of simplified models. 

A simple Lagrangian that parametrises the $T'$ couplings to quarks and electroweak bosons (showing only the couplings relevant to our analysis) is~\cite{Buchkremer:2013bha}
\begin{eqnarray}\label{Tp_lag}
\mathcal{L}_{\rm T'} &=& g^\ast \left\{ \sqrt{\frac{R_L}{1+R_L}} \frac{g}{\sqrt{2}} [ \overline{T'}_{L/R} W^+_\mu \gamma^\mu d_{L/R} ] + \sqrt{\frac{1}{1+R_L}} \frac{g}{\sqrt{2}} [ \overline{T'}_{L/R} W^+_\mu \gamma^\mu b_{L/R} ] + \right.   \label{eq:lagrangian} \\ \nonumber
&& +\left.  \sqrt{\frac{R_L}{1+R_L}} \frac{g}{2 \cos \theta_W} [ \overline{T'}_{L/R} Z_\mu \gamma^\mu u_{L/R} ] + \sqrt{\frac{1}{1+R_L}} \frac{g}{2 \cos \theta_W} [ \overline{T'}_{L/R} Z_\mu \gamma^\mu t_{L/R} ]  \right\} + h.c.\, , 
\end{eqnarray}
where the subscripts $L$ and $R$ label the chiralities of the fermions. 
Only $3$ parameters are sufficient to fully describe the interactions that are relevant for our investigation. Besides $M_{T'}$, the vector-like mass of the top partner, there are the $2$ couplings appearing in eq.~(\ref{Tp_lag}): $g^\ast$, the coupling strength to SM quarks in units of standard couplings, which is only relevant in single production (the cross sections for the latter scale with the coupling squared); and $R_L$, the generation mixing coupling, which describes the rate of decays to first generation quarks with respect to the third generation, so that $R_L = 0$ corresponds to coupling to top and bottom quarks only, while the limit $R_L = \infty$ represents coupling to first generation of quarks only.

%\section{Analysis}
All samples employed in this study have been generated up to detector level
with the \\
{\tt MadGraph5\_aMC@NLO}--{\tt PYTHIA6}--{\tt Delphes3} chain
(see details in~\cite{Basso:2014apa}).
The signal (S) is generated at leading order from the model implemented in {\tt FeynRules}. We generate $5$ benchmark points varying the $T'$ mass in steps of $200$ GeV in the range t$M_{T'} \in \left[ 800;1600\right]$ GeV,
with $g^\ast=0.1$ and $R_L = 0.5$. Contrary to the backgrounds, tau leptons have not been here included. Backgrounds (B) that can give 3 leptons in the final state which are considered in this analysis are: $t\overline{t}$ and $Z/W+jets$ with non-prompt leptonsd, and $t\overline{t}W$, $t\overline{t}Z$, $tZj$ and $VZ$ ($V=W,Z/\gamma$) with only genuinely prompt leptons. We generated leading order samples with up to 2 merged jets normalised to the (N)NLO cross section where available.

\section{Analysis}

The analysis is carried out in {\tt MadAnalysis~5}.
Leptons ($\ell = e,\, \muÂ$) and jets are required to fulfil canonical $p_T$ and $\eta$ requirements for the CMS detector.
%the following criteria: $\displaystyle
%p_T(\ell) > 20 {\mbox{ GeV,}}\, |\eta(e/\mu)| < 2.5/2.4,\,
%p_T(j) > 40 {\mbox{ GeV,}}\, |\eta(j)| < 5,\,
%\Delta R(\ell, j) > 0.4$.
External routines for $b$-tagging and for lepton isolation have been implemented. Regarding the former, here we adopted the medium working point, which has an average $b$-tagging rate of $70\%$ and a light mistag rate of $1\%$. Further, the relative isolation $I_{rel}$ is defined as the sum of the $p_T$ and calorimetric deposits of all tracks within a cone of radius $\Delta R=0.3$, divided by the $p_T$ of the lepton. The latter is isolated if $I_{rel}\leq 0.10$.
%This choice has been taken as a compromise to strongly reduce backgrounds with non-prompt leptons without suppressing the signal, where the two leptons coming from the $Z$ boson get closer and closer as the top partner mass increases.
After, we apply some general preselections as follows: we require at least 1 jet and no more than 3, of which exactly one is b-tagged, and exactly 3 leptons (electrons or muons). The requirement of less than 3 jets removes the $T'$ pair production isolating the single production channel.

%Efficiencies and event yields are evaluated for $\mathcal{L}=100$ fb$^{-1}$ and are collected in table~\ref{tab:presel_eff}. 
The requirement of $3$ isolated leptons strongly reduces the $t\overline{t}+X$ backgrounds,
%especially the $t\overline{t}+ jets$ one,
with an overall efficiency of 1 permil. The diboson component is instead strongly suppressed by the $b$-tagging, with an efficiency of $\sim 4\%$. Regarding the signal, the requirement of $3$ isolated leptons has an efficiency of $\mathcal{O}(30\%)$ and it gets less efficient as the $T'$ mass increases. This is because the $2$ leptons stemming from the $Z$ boson get closer to each other as the $T'$ gets heavier, due to the larger boost of the $Z$ boson in the $T'\to t\,Z$ decay.
Finally, the pair of same-flavour and opposite-sign leptons closest to the $Z$ boson mass is chosen, and a cut around their invariant mass distribution is performed such as $\displaystyle
|M (\ell^+ \ell^-) - M_Z| < 15 \mbox{ GeV}$. This cut removes $\sim 40\%$ $(30\%)$ of $t\overline{t}$ ($tZj$)  events.
The lepton from the top decay is therefore identified as the remaining one
% in our trilepton channel 
and labelled $\ell_W$.

We describe in the following the 2 analyses we performed, that differentiate from this point on. The first one is a traditional cut-and-count strategy, where subsequent cuts are applied to the most important kinematic variables to maximise the signal-over-background ratio. The second one is a multivariate analysis (MVA), where several discriminating observables are used at once to distinguish the signal from the background, cutting at the end only on its output.

%%%%%%%%%%%%%%%%%%%%%%%%%%%%%%%%%%%%%%%%%%%%%%%%%%%%%%%%%
%			cut-and-count			%
%%%%%%%%%%%%%%%%%%%%%%%%%%%%%%%%%%%%%%%%%%%%%%%%%%%%%%%%%

\begin{SCfigure}
\centering
\includegraphics[width=0.7\linewidth]{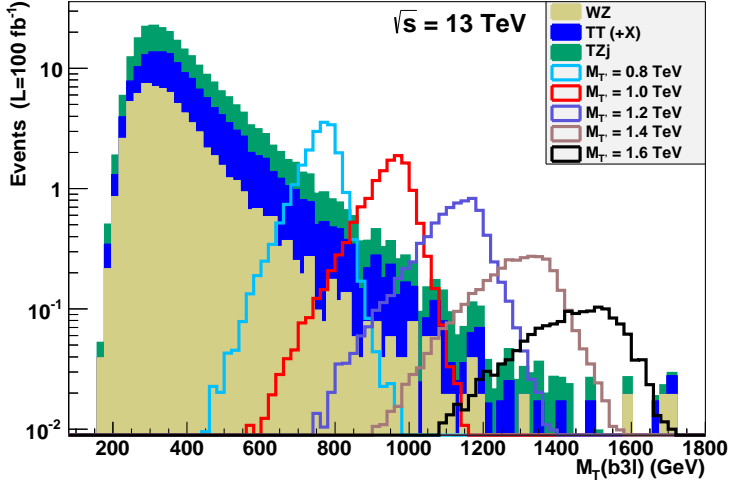} 
\caption{Transverse mass distribution for the $T'$ decay products: the $3$ charged leptons and the $b$-jet.}
\label{fig:MtZ}
\end{SCfigure}

%\subsection{Cut-and-count}\label{sect:cc}
The first strategy to study the LHC discovery potential illustrated here is the cut-and-count one (C\&C).
The $W$ boson and the top quark are reconstructed as resonances in the transverse mass distributions of the decay products, here chosen
because of the sharper peaks as compared to those employing the invariant mass. We apply loose selections as follows:
$\displaystyle 10 < M_T (\ell_W \nu) / \mbox{GeV} < 150$ and
$0 < M_T (\ell_W\, b \nu) / \mbox{GeV} < 220$.
In particular, the lower cut for $M_T (\ell_W \nu)$ is inspired by experimental analyses to suppress the multijet background, which we did not simulate.
%Surviving events and relative efficiencies are collected in table~\ref{tab:cc_eff}.
 These numerical values have been chosen to maximise the signal-over-background ratio while keeping at least $90\%$ of the signal. For the backgrounds, the top-mass reconstruction has an efficiency of $\sim 60\%$ $(50\%)$ for $t\overline{t}$ ($WZ$).
Contrary to ref.~\cite{expVLQ}, we do not require a forward jet to not suppress any further the signal, despite it being a distinctive feature of our signature. This is also not necessary: the signal is already clearly visible above the background in the distribution of the transverse mass of the $T'$ decay products (the $3$ charged leptons and the $b$-jet), as can been seen in figure~\ref{fig:MtZ}.
% for the various signal benchmark points.

%We select a window around the peak of each benchmark point to get the best signal-over-background significance and we collect the final numbers in table~\ref{tab:significances}.

%%%%%%%%%%%%%%%%%%%%%%%%%%%%%%%%%%%%%%%%%%%%%%%%%%%%%%%%%
%			MVA				%
%%%%%%%%%%%%%%%%%%%%%%%%%%%%%%%%%%%%%%%%%%%%%%%%%%%%%%%%%

%\subsection{Multivariate analysis}\label{sect:MVA} 
%In section~\ref{sect:cc} we
The analysis just presented showed that suitable cuts on the most straightforward distributions were sufficient to isolate the signal from the background.
One could wonder if this was the best strategy, i.e. cutting on those variables with the values we chose.
There are in fact many additional variables that one could analyse to distinguish the signal from the background. However, cutting on any of these variables will unavoidably reduce also the signal. To overcome this,
several variables can be combined using a multivariate analysis (MVA)
to obtain the best signal/background discrimination. We identified some discriminating variables in table~\ref{tab:MVA_var}, ranked according to their discriminating power when a boosted decision tree (BDT) is employed. Here, $\Delta\varphi$ is the difference of the azimuthal angles between 2 objects, $\Delta\eta$ is the difference of their pseudorapidities, and $\Delta R = \sqrt{(\Delta\varphi)^2 + (\Delta\eta)^2}$. Further, the presence of a forward jet is a prominent feature of the signal. To account for this, we use the largest pseudorapidity of all jets $\eta^{\mbox{max}}(j)$ in the event.

\begin{SCtable}
\centering
\scalebox{0.84}{
 \begin{tabular}{|c|c|c|c|}
 \hline
 Variable  & Importance &  Variable  & Importance   \\
 \hline
$M_T(b\,3\ell)$ & $2.60\, 10^{-1}$ & $\Delta R (b,\, \ell_W)$ & $9.77\, 10^{-2}$ \\
$p_T(Z) / M_T(b\,3\ell)$ & $9.41\, 10^{-2}$ & $\Delta\varphi (t,\, Z)$ & $8.17\, 10^{-2}$ \\
$\eta^{\mbox{max}} (j)$ & $6.02\, 10^{-2}$ & $\Delta\varphi (\ell\ell|_Z)$ & $5.89\, 10^{-2}$\\
$\Delta\varphi (Z,\, \slashed{p}_T)$ & $5.37\, 10^{-2}$ & $p_T(j_1) / M_T(b\,3\ell)$ & $5.08\, 10^{-2}$\\
$\Delta\eta (\ell\ell|_Z)$ & $5.05\, 10^{-2}$ & $\Delta\eta (b,\, \ell_W)$ & $5.03\, 10^{-2}$ \\
$\eta(t)$ & $4.99\, 10^{-2}$ & $\Delta\varphi (Z,\, \ell_W)$ & $4.63\, 10^{-2}$\\
$\eta(Z)$ & $4.61\, 10^{-2}$ &  & \\ \hline
 \end{tabular}}
 \caption{Ranking training variables for $M_{T'}=1.0$ TeV and full background. Here $\ell\ell|_Z$ identifies the 2 leptons that reconstruct the $Z$ boson.}
 \label{tab:MVA_var}
\end{SCtable}

Trivial correlations (such as between the $T'$ mass, the $p_T$ of the leading jet and the $p_T$ of the $Z$ boson) are efficiently removed if one consider ratios of those $p_T$'s over $M_T(b\,3\ell)$.
All other variables are almost uncorrelated, with a degree of correlation of $\pm30\%$ at most. We also checked that the MVA does not suffer of overtraining.
The variables in table~\ref{tab:MVA_var} are used to train the BDT to recognise the signal against the background. They are selected after the $Z$ mass reconstruction. The BDT trained on each benchmark point is then applied on the full signal and background samples.

\subsection{Results}\label{sect:results}
We collect here the final results for the discovery power at the LHC. 
In the case of the cut-and-count analysis, 
we need to select a window around the signal peaks in the $M_T(b 3\ell)$ distribution. For the MVA analysis,
we need to perform a cut on the BDT output that maximises the significance. The maximum significance for the benchmark points, evaluated as $\sigma = S/\sqrt{S+B}$, are collected in table~\ref{tab:significances}.

\TABLE{
\centering
\scalebox{0.84}{
 \begin{tabular}{|cc|c|c|c|c|c|}
 \hline
\multicolumn{2}{|c|}{Analysis} & $M_{T'}=0.8$ TeV & $M_{T'}=1.0$ TeV & $M_{T'}=1.2$ TeV & $M_{T'}=1.4$ TeV & $M_{T'}=1.6$ TeV\\
 \hline
\multicolumn{2}{|c|}{$M_T(b3\ell)$ cut (GeV)} & $[800-860]$    & $[840-1200]$    & $[1000-1340]$   & $[1120-1640]$ & $[1200-1800]$ \\
\multirow{3}{*}{C\&C} &S (ev.)  &18.00  &12.28  & 7.16  & 3.40  & 1.57 \\
		      &B (ev.)  & 8.90  & 4.88  & 1.74  & 0.90  & 0.63 \\
		      &$\sigma$ & 3.47  & 2.96  & 2.40  & 1.64  & 1.06 \\ \hline
\multirow{2}{*}{MVA} & cut   & 0.07 & 0.08 & 0.11 & 0.12 & 0.12 \\
 & $\sigma$                  & 3.64 & 3.10 & 2.50 & 1.62 & 1.15 \\  
 \hline
 \end{tabular}}
 \caption{Signal and background events and maximum significance for the benchmark points for $\mathcal{L}=100$ fb$^{-1}$, after selecting a mass window (for the C\&C), or after cutting on the BDT output (MVA).}
 \label{tab:significances}
}

One of the most important results here described is that the dedicated BDT analysis does not significantly improve on the cut-and-count strategy.
The latter analysis is certainly sufficient and easier. The cuts as above described
are already best optimised, as is the signal peak selection. No further variable/cut need to be considered/applied.

The significances in table~\ref{tab:significances} are for the benchmark points.
%, evaluated for $g^\ast = 0.1$ and $R_L=0.5$. 
We can now extrapolate them to the full $g^\ast$--$R_L$ parameter space. The $3$ and $5$ sigma discovery lines are drawn as a function of $g^\ast$ and the $T'$ mass for some fixed values of $R_L$ in figure~\ref{fig:significance_100}(left), and as a function of $g^\ast$ and $R_L$ for the benchmark $T'$ masses in figure~\ref{fig:significance_100}(right).
These plots show that with $100$ fb$^{-1}$ of data, $T'$ masses up to $2$ TeV can be observed. The cross section for the trilepton decay channel of the $T'$ (and hence the LHC reach) increases considerably when $R_L$ is non-vanishing, getting to a maximum for $R_L\simeq 1$, corresponding to $50\%$--$50\%$ mixing. This effect is simply due to the increased admixture of valence quarks in production, mitigated by a reduced $T'$-to-$tZ$ branching ratio, as $R_L$ increases.
The reach in $g^\ast$ is here roughly twice than for the no mixing case ($R_L=0$).

\FIGURE{
\centering
\includegraphics[width=0.45\linewidth]{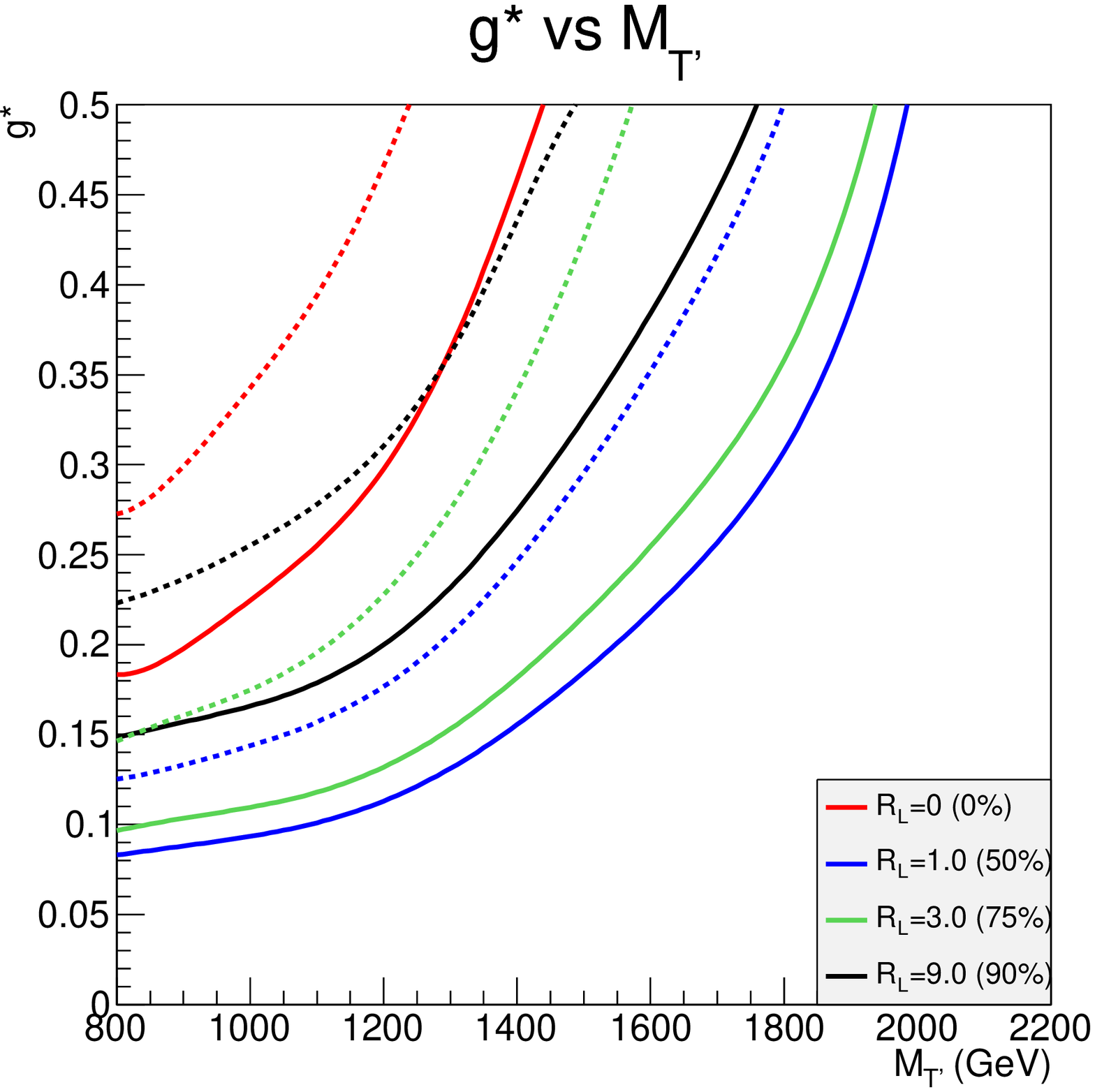} 
\includegraphics[width=0.45\linewidth]{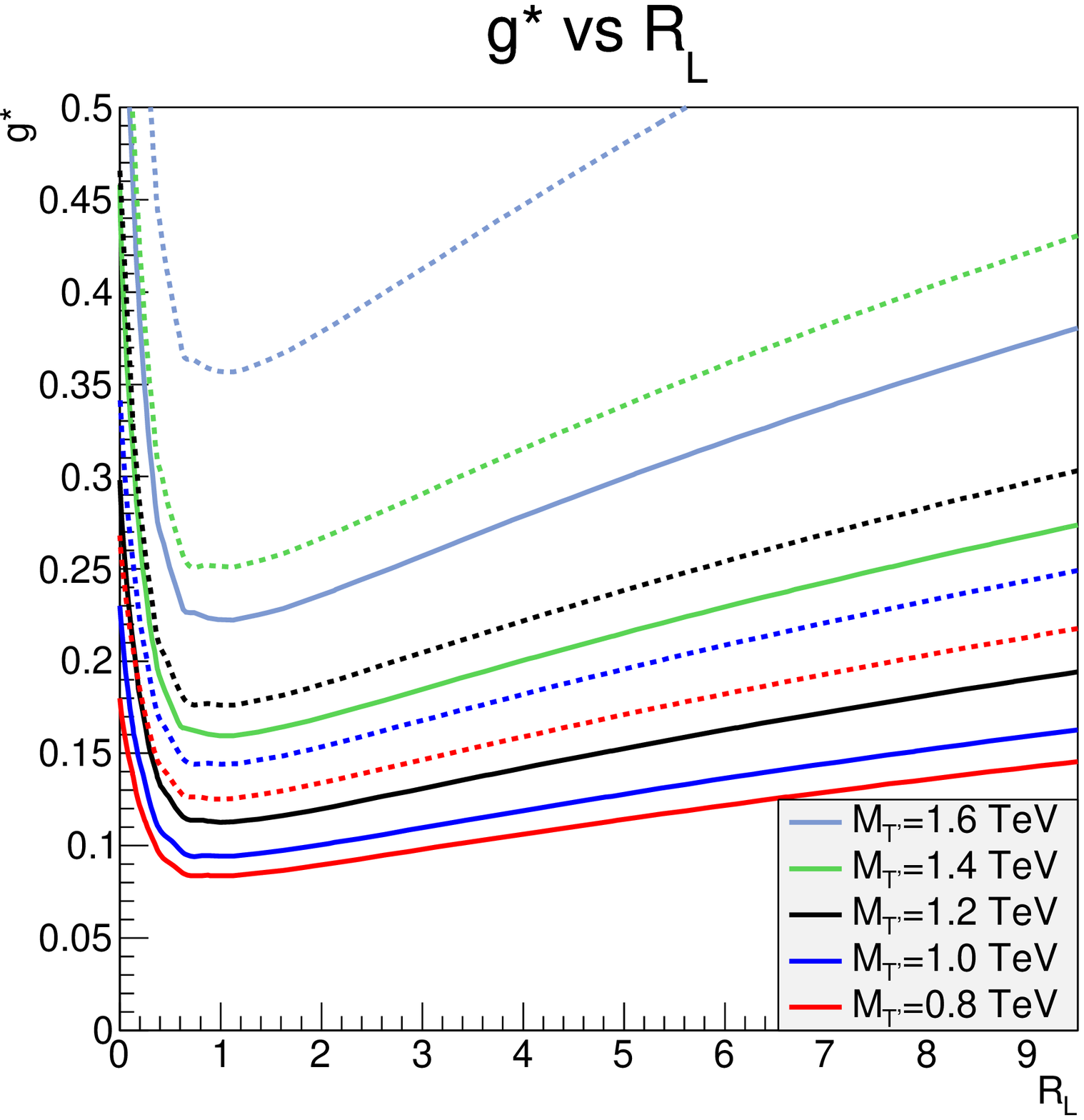} 
\caption{Significance $\sigma=3$ (solid lines) and $\sigma=5$ (dashed lines) for $\mathcal{L}=100$ fb$^{-1}$.}
\label{fig:significance_100}
}

\subsection{Top FCNC reinterpretation}\label{sect:FCNC}
We conclude by presenting a reinterpretation of our investigation in terms of the top-quark FCNC coupling to a light quark and a $Z$ boson. In this scenario, the top quark interacts with a $Z$ boson and a up- or charm-quark via the $\kappa_{tZq}$ coupling appearing in the FCNC Lagrangian~\cite{AguilarSaavedra:2008zc}
$\displaystyle
\mathcal{L}= \sum_{q=u,c} \frac{g}{\sqrt{2}c_W} \frac{\kappa_{tZq}}{\Lambda}\,\overline{t}\sigma^{\mu\nu}\,\big(f^L_{Zq} P_L + f^R_{Zq} P_R \big)\,q\, Z_{\mu\nu}\, ,
$
where $\Lambda$ is the scale of new physics.
This Lagrangian gives a similar final state as the one here described, $pp\to tZ$, with a top-quark and a $Z$ boson produced back-to-back.
The analyses of the $T'$-mediated signature subject of this paper could therefore be as well sensitive to the one induced by the top effective coupling. We tested it by producing at leading order $pp\to tZ$ samples when turning on one FCNC coupling at the time, labelled $\kappa_{tZu}$ and $\kappa_{tZc}$, respectively,
that have been analysed following the cut-and-count strategy. 

\begin{comment}
Efficiencies and event yields for $100$~fb$^{-1}$ are collected in table~\ref{tab:fcnc_eff}.

\begin{SCtable}
\centering
 \begin{tabular}{|c|c|c|}
 \hline
 Cut  & $\kappa_{tZu}$  & $\kappa_{tZc}$ \\
 \hline
 no cuts &  $2263(100\%)$ & $5360(100\%)$\\
 $1\leq n_j\leq 3$ & 1765(78.0\%) & 4452(83.0\%)  \\
 $n_\ell \equiv 3$ & 191.8(10.9\%) & 623.3(14.0\%) \\
 $n_b \equiv 1$ & 113.8(59.3\%)	 & 381.0(61.1\%) \\
 $Z$-reco   & 103.2(90.7\%)  & 342.7(90.0\%) \\
 $W$-reco   & 96.2(93.3\%)   & 323.6(94.4\%)\\
 $t$-reco & 91.1(94.7)     & 304.7(94.1\%)\\ \hline
$M_T(b3\ell)$ & $>400$ GeV & $>200$ GeV \\ \hline
 S & 68.0  & 304.5\\
 B & 102.9 & 241.7\\
 $\sigma$ & 5.2 & 13.0\\ \hline
 \end{tabular}
 \caption{Surviving events (and efficiencies with respect to the previous item) for the cut-and-count analysis of the FCNC top coupling $\kappa_{tZu}/\Lambda$ (at current limit) and $\kappa_{tZc}/\Lambda$ (for BR$(t\to Zc)=1\%$), and signal/background events that maximise the significance.}
 \label{tab:fcnc_eff}
\end{SCtable}
\end{comment}

The significance for the $\kappa_{tZu}$ sample is maximised by selecting $M_T(b 3\ell) > 400$ GeV, reaching the value of $5.2$ sigma for the present best limit of the coupling of $\kappa_{tZu}/\Lambda= 0.2$~TeV$^{-1}$ (or BR$(t\to Zu)= 0.05\%$)~\cite{Chatrchyan:2013nwa}, corresponding to a cutoff scale $\Lambda=5$ TeV. Regarding the $\kappa_{tZc}$ sample, we chose a coupling yielding BR($t\to Zc)=1\%$ to compare the results. For this value, the highest significance of $13.0\sigma$ is obtained by selecting $M_T(b 3\ell) > 200$ GeV.
The MVA trained on each $T'$ signal has been applied to the FCNC samples but, also in this case, it did not improve the sensitivity.

\section{Conclusions}\label{sect:conclusions}
In this work we described the LHC Run-II discovery potential of the trilepton channel for a singlet top partner in the single production mode and its subsequent decay into a top quark and a $Z$ boson. A simple cut-and-count analysis has been designed, by selecting and cutting the most straightforward distributions. A suitable multivariate analysis did not improve significantly on the cut-and-count results. The comparison was performed on several signal benchmark points.

Overall, a search at the LHC in the trilepton channel can be sensitive to top partners decaying into $tZ$ for masses up to $2.0(2.1)$ TeV and couplings down to $0.08(0.05)$ with $100(300)$ fb$^{-1}$ of data. 
Finally, we reinterpreted our analyses in the context of a top FCNC coupling to a $Z$ boson and a light quark, which provides a similar final state. We showed that this channel can discover at $5\sigma$ values of the couplings at the present best exclusion limit (for $100$~fb$^{-1}$), probe at $3\sigma$ FCNC branching ratios down to $0.025\%(0.16\%)$ for $\kappa_{tZu}/\Lambda$($\kappa_{tZc}/\Lambda$), or eventually extend the exclusion limits down to 
$0.016\%$ and $0.1\%$ for the two FCNC couplings, respectively.

%\section*{Acknowledgements}
%We would like to sincerely thank our colleagues A.~Alloul, C.~Collard, E.~Conte and G.~Hammad for the help in generating the background samples used in this work and for useful comments. The work of LB is supported by the Theorie-LHC France initiative of the CNRS/IN2P3 and by the French ANR 12 JS05 002 01 BATS@LHC.

\end{document}